\documentclass[prb,twocolumn,aps,showpacs,fixfloats]{revtex4}
\usepackage{graphicx}
\usepackage{bm}
\usepackage{amsmath,amssymb}
\usepackage{subfigure}
\usepackage{float}
\usepackage{latexsym}
\usepackage{epstopdf}
\usepackage{color}
\usepackage{enumerate}
\usepackage{pdfpages}

\newcommand{\s}{\scriptscriptstyle}

\newcommand{\R}{\mathcal{R}}
\newcommand{\Rs}{\widetilde{\mathcal{R}}}

\begin{document}

\title {Spin transport in n-type  single-layer transition metal dichalcogenides }

\author{Z. Yue$^{1}$, Kun Tian$^{2}$, A. Tiwari$^{2}$,  and M. E. Raikh$^{1}$ }

 \affiliation{$^{1}$Department of Physics and
Astronomy, University of Utah, Salt Lake City, UT 84112, USA \\
$^{2}$Department of Materials Science and Engineering, University of Utah, Salt Lake City, Utah 84112, USA
}



\begin{abstract}
Valley asymmetry of the electron spectrum in transition metal dichalcogenides (TMDs) originates from the spin-orbit
coupling. Presence of spin-orbit fields of opposite signs for electrons in $K$ and $K'$ valleys in combination with
possibility of intervalley scattering result in a nontrivial spin dynamics. This dynamics is reflected in the dependence
of nonlocal resistance on external magnetic field (the Hanle curve). We calculate theoretically the Hanle shape in TMDs.
It appears that, unlike conventional materials without valley asymmetry, the Hanle shape in TMDs is different for normal and parallel orientations of the external field. For normal orientation, it has two peaks for slow intervalley scattering,
while, for fast intervalley scattering the shape is usual. For parallel orientation, the Hanle curve exhibits a cusp at zero field. This cusp is a signature of a slow-decaying valley-asymmetric mode of the spin dynamics.
\end{abstract}

\maketitle

\begin{figure}
\includegraphics[width=84mm]{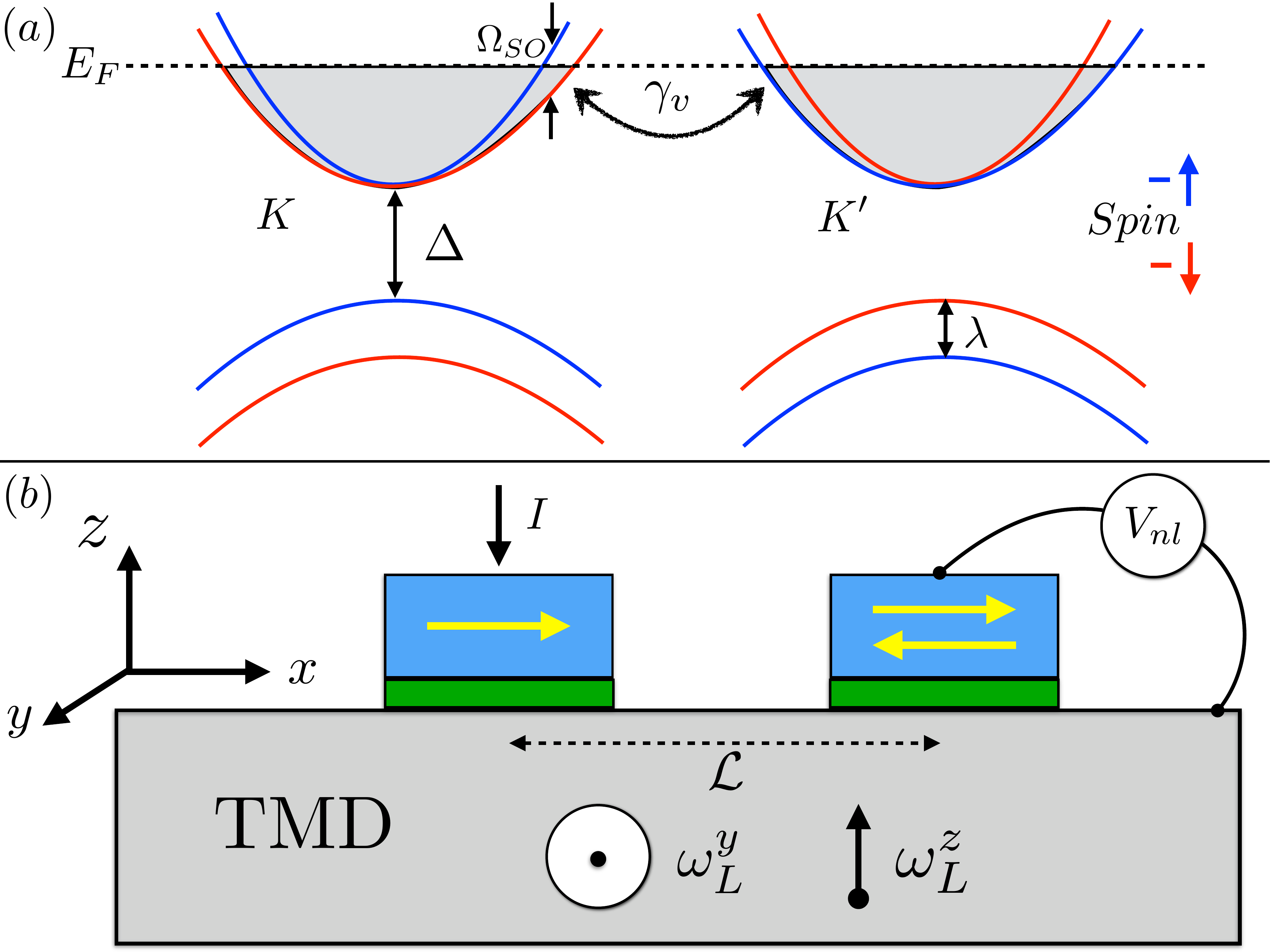}
\caption{(Color online) (a) The energy spectrum of TMD at $K$ and $K'$ valleys is shown schematically. In n-type material the states with energies below $E_F$ are occupied. The splitting between the $\uparrow$ and $\downarrow$ branches at the Fermi level is $\Omega_{\s SO}=\big(\frac{\lambda}{\Delta}\big)E_F$ and is much smaller than $E_F$. Short-range impurities are responsible for intervalley scattering with a rate $\gamma_v$. (b) Schematic illustration of the spin-transport experiment in TMD. The measure of the nonlocal resistance is a voltage
between the channel and the right ferromagnet detector generated upon injecting the current
through the left ferromagnetic electrode.  Injector and detector shown in blue are separated from the channel by a tunnel barrier shown in green. While the injected polarized electrons travel
either in the valley $K$ or in the valley $K'$,  their spin precesses in the effective field ${\bm \omega}_L+{\hat z}\Omega_{\s SO}$
or ${\bm \omega}_L-{\hat z}\Omega_{\s SO}$. As a result of intervalley scattering, the spin dynamics is described by
a system of the coupled equations Eq. (\ref{general}). We study the limit of a small, compared to the diffusion length, distance ${\cal L}$ between  injector and detector. The specifics of TMDs is that for external field, ${\bm \omega}_L$, oriented along $z$ and $y$  the shapes of the Hanle curves $V_{nl}(\omega_L)$ are completely different.}
\label{figure1}
\end{figure}

\section{Introduction}

 Transition-metal dichalcogenides (TMDs) in a 2D domain are single-layer semiconductors with
lattice similar to graphene.
Unlike graphene, they possess a bandgap, which makes them attractive for optoelectronic applications, such as field-effect transistor\cite{Transistor1,Transistor2}, see e.g. Ref. \onlinecite{Review} for review.      Unlike graphene,  the electron states in  $K$ and $K'$ valleys are not equivalent. This inequivalence is owed to the spin-orbit coupling. The $K$ and $K'$ wave functions corresponding to the same momenta and
energies differ by the spin direction. Spin-orbit splitting of the conduction band is
much smaller than that of the valence band.\cite{KP1,KP2,KP3,KP4,KP5,KP6,KP7} As a result of the band splitting,
there are two excitons in a given valley.
Correspondingly, in undoped samples, the spectra of the exciton absorption, reflection, and luminescence
exhibit a two-peak structure,
as was demonstrated experimentally by many groups\cite{OpticalExperimentHeinz,OpticalExperimentNatureNano,
OpticalExperimentNaturePhysics,OpticalHanleNatureCobden,
ClassicalHeinz1,ClassicalHeinz2,Biexciton,IntervalleyPump,Glazov2,Glazov3}

Upon photoexcitation of  n-type samples, generated holes rapidly recombine with resident electrons, while generated electrons preserve  spin memory for  rather long times  $\sim 1$ns. This was
established in Refs. \onlinecite{Crooker1,Crooker3} on the basis of the analysis of the Hanle-Kerr data is magnetic field parallel to the layer. The fact that the optical response is sensitive to a magnetic field of $\sim 50$mT which is much smaller than the effective spin-orbit (SO) field seems rather unusual. The
explanation for that suggested in Ref. \onlinecite{Crooker1} is based on the fact that 
an electron, created by light, undergoes fast inter-valley scattering, which effectively
averages out the SO field. This scattering is facilitated by disorder, unlike the intervalley scattering of 
excitons\cite{Glazov1,Wu1} which is facilitated by the exchange interaction. The latter mechanism is  
similar to the non-radiative F{\"o}rster energy transfer.
Optical response to a magnetic field perpendicular to the layer emerges only when the field is very strong\cite{Crooker2} $\sim 50$T.

Gate voltages\cite{Gate} can control the type and the concentration of carriers in TMDs.  However, the transport measurements reported to date are scarce compared to the optical studies. The highest mobility reported to date\cite{HighMobilityNatureLett} in n-type MoS$_2$ is $\sim 10^3$cm$^2$/Vs.
For most samples the mobility is lower\cite{HighMobilityNature} $\sim 10^2$cm$^2$/Vs.
 The fact that it depends on temperature\cite{HighMobilityNatureLett,HighMobilityNature} suggests that the electron states are not far from the metal-insulator transition. Hopping transport has also been reported
 in disordered MoS$_2$ samples.\cite{Hopping}

Spin transport has never been studied in TMDs\cite{footnote}. On the other hand, relatively low mobility does
not prevent such studies, see e.g. Ref. \onlinecite{TiwariZnO}. Note that the spin transport and
the Kerr rotation signals  
are both limited by the spin-memory loss of carriers. In this regard, the most interesting question is how the spin dynamics of electrons reflected in  the spin transport  in n-type TMDs is related to the spin dynamics
of excitons inferred from the Hanle-Kerr measurements\cite{Crooker1,Crooker3}. This issue
is studied theoretically in the present paper. One cannot expect an observable spin transport in p-type TMDs. The separation of $\sim 150$meV between the tops of
  $\uparrow$ and $\downarrow$ bands in each valley suggests that ``intrinsic"  spin precession is too fast. Intervalley scattering is also strictly forbidden unless phonons are involved\cite{Dery1,Dery2}.

There are apparent differences between the spin-transport studies and polarization-of-luminescence techniques.
Firstly, the optical experiments reveal the dynamics of the $z$-component of spin, $S_z(t)$, while conventional spin transport measures $S_x(t)$, $S_y(t)$, as illustrated in the Fig. \ref{figure1}.
Secondly, the magnitude of the SO field in the metallic regime depends
strongly on the electron density and is much smaller than for the excitons.
It is also nontrivial that the electron inter-valley scattering rate, $\gamma_{v}$, depends on
the concentration of the short-range impurities (defects\cite{Defects}) allowing the large momentum transfer
between the valleys. Finally, the F{\"o}rster-like mechanism, which is at work for excitons, does not apply for
two Fermi seas at $K$ and $K'$ valleys. As for relation between the spin transport and the Kerr rotation techniques,
the latter also studies $S_z(t)$. Besides, the Kerr rotation signal is pronounced for probe frequencies near the  A-exciton resonance\cite{Crooker1,Crooker3} not at the Fermi level.

We will demonstrate that the shapes of the transport Hanle
curves in TMDs depend dramatically on the ratio of $\gamma_{v}$ to the SO splitting of the electron
spectrum, $\Omega_{\s SO}$, and that these shapes  are different  from the conventional
transport  Hanle curves. In this regard, we emphasize, that the shapes of the Hanle curves reported for wide variety of materials are very robust\cite{Robust}.
Specifics  of the Hanle curves in TMDs is due to the valley asymmetry.

We show that, unlike conventional materials,  the shape of the transport Hanle curve in TMDs depends on the orientation of the external field. If the spin polarization of electrons injected from a ferromagnet is along the $x$-axis, see Fig. \ref{figure1}, the dynamics of the injected spin is different for the external field
parallel to the layer (along $y$) or normal to the layer (along $z$).
For normal orientation and $\gamma_{ v}\ll \Omega_{\s SO}$ this dynamics is different for different valleys. As a result, of this the Hanle curve has a two-peak structure.
A distinct Hanle shape also persists for the normal orientation when $\gamma_{ v}\gg \Omega_{\s SO}$. It
represents the difference of  two conventional Hanle profiles with very different widths: the wider reflecting the valley-symmetric mode of the overdamped spin dynamics, while the narrower reflecting the valley-antisymmetric mode.

When the external field is parallel to the layer, the Hanle curve has a singularity at a zero field. This singularity
is due to the valley-asymmetric mode describing the slow time decay of the spin density.
The decay is slow as a result of fast alternation of valleys; external field is responsible for coupling
of the initial spin distribution to this mode. Characteristic width of the Hanle curve for the parallel orientation of external field is much smaller than for normal orientation. This is in accord with results  experimental findings\cite{Crooker1,Crooker3,Crooker2} for magnetic-field response of photoexcited carriers,
and is not surprising, since the spin dynamics for electrons and excitons are qualitatively similar.

\section{Density dependence of the SO splitting of the electron spectrum}

The ${\bm k}\cdot{\bm p}$ Hamiltonian of a TMD, established in Ref. \onlinecite{Hamiltonian}, see also Refs. \onlinecite{KP1,KP2,KP3,KP4,KP5,KP6,KP7}, contains three energies, namely, the gap, $\Delta$,
the hopping integral, $t$, and the SO-induced spin splitting of the valence band top,
$2\lambda$. With two valleys coupled to two spin  projections, it represents a $4\times 4$ matrix. In the presence of external field having $y$ and $z$ components,
this matrix has the form

 \begin{equation}
\hspace{-4mm}H\hspace{-1mm}=\hspace{-1mm}
\begin{pmatrix}
\frac{\Delta}{2}+\omega_L^{z} & at\tau ke^{-i\theta \tau} & -i\omega_L^{y} & 0 \\
at\tau ke^{i\theta \tau}  & -\frac{\Delta}{2}+\omega_L^{ z}+\lambda \tau & 0 &   -i\omega_L^{ y} \\
i\omega_L^{y} & 0 & \frac{\Delta}{2}-\omega_L^{z} &  at\tau ke^{-i\theta \tau}\\
0 &  i\omega_L^{ y} &  at\tau ke^{i\theta \tau} & -\frac{\Delta}{2}-\omega_L^{z}-\lambda \tau
\end{pmatrix},
\end{equation}
where $a$ is the lattice constant, $\omega_L^{y}$ and $\omega_L^{ z}$ are the corresponding Zeeman energies,
$k$ and $\theta$ are the magnitude and the orientation of the wave vector. The valley index $\tau$ takes the values $\pm 1$.

The spectrum, $\varepsilon(k)$, originating from the Hamiltonian Eq. (3) is the solution of the fourth-order equation
\begin{align}
&\Big[\big( \varepsilon+\frac{\Delta}{2}+\omega_L^{ z}+\lambda \tau \big)\big(\varepsilon-\frac{\Delta}{2}+\omega_L^{ z}\big)-(atk)^2\Big] \nonumber \\
\times &\Big[\big( \varepsilon+\frac{\Delta}{2}-\omega_L^{ z}-\lambda \tau \big)\big(\varepsilon-\frac{\Delta}{2}-\omega_L^{ z}\big)-(atk)^2\Big]   \nonumber \\
=&\Big[2 \varepsilon^2+2\big(\frac{\Delta}{2}\big)^2+2(atk)^2-(\omega_L^{ z})^2-(\omega_L^{ z}+\lambda \tau)^2\Big](\omega_L^{y})^2.
\end{align}
In the absence of magnetic field the right-hand side is zero, and each bracket in the
left-hand side determines the corresponding branch of the spectrum. With magnetic field, we can find the spectrum of the conduction band perturbatively in the small parameter $\omega_L^{ y}/\Delta$. The result reads
\begin{equation}
\label{spectrum}
\varepsilon(k)=\frac{\Delta}{2}+\frac{\hbar k^2}{2m_c}\pm \sqrt{\Big[\omega_L^{ z}+\Big(\frac{\lambda \tau}{\Delta}\Big)\frac{\hbar k^2}{2m_c}\Big]^2+(\omega_L^{ y})^2}.
\end{equation}
Here $m_c=\Delta\hbar/2a^2t^2$ is the effective mass of the conduction-band electron.
Relative splitting of $\uparrow$ and $\downarrow$ branches  is always small by virtue of the
parameter $\lambda/\Delta$, which is $\approx 0.1$ for MoS$_2$.
The above result has a simple interpretation. Namely, $\Omega_{\s SO}$ acts as an effective field
directed along $z$ which assumes opposite values for two the valleys.

In n-type TMDs the electron states  with  $k<k_{\s F}$, where $k_{\s F}$ is the Fermi momentum, are occupied. The parameter crucial for spin transport is the ratio, $\gamma_v/\Omega_{\s SO}$, of the intervalley scattering rate and the band splitting,\cite{footnote2}
$\Omega_{\s SO}=\left(\frac{\lambda}{\Delta}\right)E_{\s F}$, at the Fermi level
$E_{\s F}=\hbar k_{\s F}^2/2m_{c}$. We can perform a numerical estimate of this ratio
assuming that the mobility is limited by the same short-range impurities that are
responsible for intervalley  scattering. The fact that point-like defects are the
leading source of scattering in TMDs is commonly accepted, see e.g. Ref. \onlinecite{Defects}. With mobility given by $\mu=\frac{e}{m_c\gamma_v}$ and
$k_{\s F}^2=4\pi n$, where $n$ is the electron density, we find
\begin{equation}
\label{ratio}
\Gamma=\frac{\gamma_v}{\Omega_{\s SO}}=\left(\frac{\Delta}{\lambda}\right)\frac{e}{2\pi\hbar\mu n}.
\end{equation}
For numerical estimate we choose a typical value $n=10^{13}cm^{-2}$. Then for the highest reported mobility\cite{HighMobilityNatureLett} for electrons in MoS$_2$,  $\mu=10^3 cm^2/Vs$, the ratio Eq. (\ref{ratio}) is equal to $0.2$, while
for typical mobility\cite{HighMobilityNature} $\mu=10^2 cm^2/Vs$ it is $10$ times bigger. Thus, we
conclude that both regimes $\Gamma \ll 1$ and $\Gamma \gg 1$ are viable for spin transport.

\section{Nonlocal resistance}
Once the spectrum Eq. (\ref{spectrum}) in magnetic field is known and the intervalley scattering rate
is introduced, the procedure of calculation of  nonlocal resistance is straightforward. \cite{vanWeesPioneering} First the splitting of the spectrum is incorporated into the equation of the dynamics for the
spin density ${\bm S}(t)$ which is solved with an initial condition  ${\bm S}(0)={\hat x}$. Then the solution for $S_x(t)$ is multiplied by the diffusion propagator
\begin{equation}
\label{propagator}
P_{{\cal L}}(t)=\frac{1}{\left(4\pi Dt\right)^{1/2}}\exp\left(-\frac{{\cal L}^2}{4Dt}\right),
\end{equation}
where ${\cal L}$ is the distance between the injector and detector and $D$ is the diffusion coefficient related
to mobility via the Einstein relation. Finally, the nonlocal resistance is obtained by integration over
time
\begin{equation}
\label{R}
R(\omega_L) = {\cal R}_0 \int\limits_0^\infty dt S_x(t) P_{{\cal L}}(t),
\end{equation}
where ${\cal R}_0$ is the prefactor.
The specifics of TMDs is that $S_x(t)$ is the sum $S_x(t)=S_x^K(t)+S_x^{K'}(t)$ of contributions
of the two inequivalent valleys, so that the spin dynamics is governed by the system of the coupled equations\cite{Crooker1,footnote1}
\begin{align}
\label{general}
\frac{d{\bm S}^K}{d t} &= \omega_L^{y} {\hat y} \times {\bm S}^K + (\Omega_{\s SO}+ \omega_L^{z}){\hat z} \times  {\bm S}^K
\hspace{-2mm} - \gamma_v \left( {\bm S}^K\hspace{-1.2mm} -\hspace{-0.7mm} {\bm S}^{K'} \right), \nonumber \\
\frac{d {\bm S}^{K'}}{d t} &=  \omega_L^{ y} {\hat y} \times {\bm S}^{K'} \hspace{-2mm}- (\Omega_{\s SO}- \omega_L^{ z}){\hat z} \times  {\bm S}^{K'}
\hspace{-2mm}+\gamma_v \left( {\bm S}^K\hspace{-1.2mm} - \hspace{-0.7mm}{\bm S}^{K'} \right).
\end{align}
We will consider the cases of the normal, $\omega_L \parallel z$,  and
tangential, $\omega_L \parallel y$, orientations of the external field separately.

\begin{figure}
\includegraphics[width=84mm]{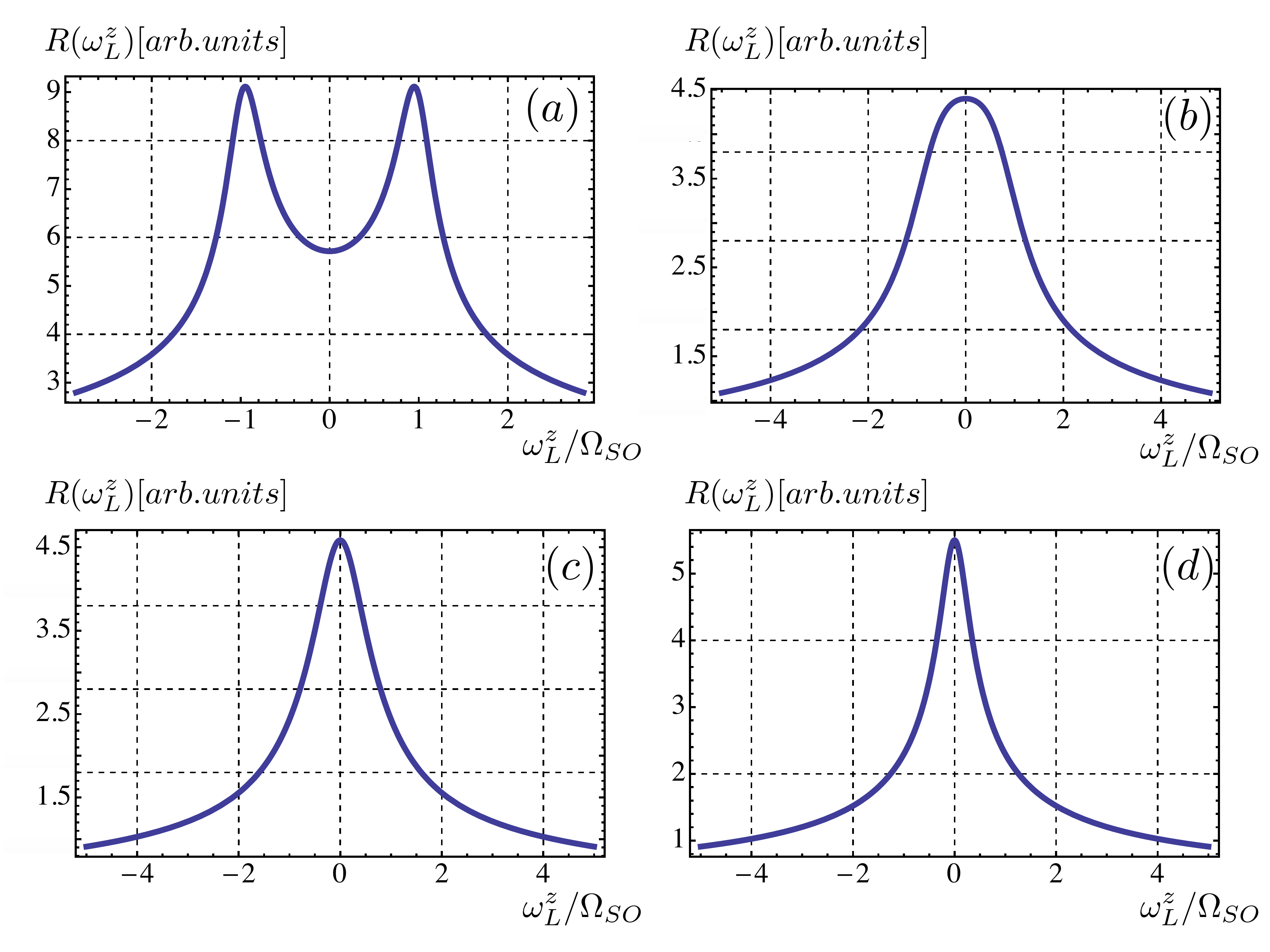}
\caption{(Color online) Nonlocal resistance  calculated from Eqs. (\ref{Gamma<1}), (\ref{Gamma>1}) is plotted versus the dimensionless
magnetic field normal to the plane for different intervalley scattering rates (in the units of $\Omega_{\s SO}$): $\Gamma=0.2$ (a), $\Gamma=0.7$ (b), $\Gamma=1.2$ (c), and $\Gamma=1.8$ (d). Two-peak structure of the Hanle curves centered  at $\omega_L^z=\pm \Omega_{\s SO}$ evolves with increasing $\Gamma$ to a conventional Hanle shape.}
\label{figure2}
\end{figure}

\section{Normal orientation of the external field}
For normal orientation, the $z$-component of the spin drops out from the system
Eq. (\ref{general}). To analyze this system, it is convenient, following Ref. \onlinecite{Crooker1}, to introduce, in addition to the net spin projections $S_x(t)$ and $S_y(t)$,
the valley imbalances
\begin{equation}
 S_x^{-}=S_x^K-S_x^{K'},~~~~ S_y^{-}=S_y^K-S_y^{K'}.
\end{equation}
Upon the Laplace transform, the system of four equations for $S_x$, $S_y$, $S_x^{-}$, and $S_y^{-}$
assumes the form
\begin{align}
\label{system1}
p{\tilde S}_x-1 &= -\Omega_{\s SO}{\tilde S}_y^- -\omega_L^z{\tilde S}_y \nonumber \\
p{\tilde S}_y &= \Omega_{\s SO}{\tilde S}_x^- +\omega_L^z{\tilde S}_x \nonumber \\
p_1{\tilde S}_x^- &= -\Omega_{\s SO}{\tilde S}_y -\omega_L^z{\tilde S}_y^- \nonumber \\
p_1{\tilde S}_y^- &= \Omega_{\s SO}{\tilde S}_x +\omega_L^z{\tilde S}_x^- ,
\end{align}
where ${\tilde S}(p)$ stands for the Laplace-transformed $S(t)$, and $p_1$ is defined as
\begin{equation}
p_1=p+2\gamma_v.
\end{equation}
The solution of the system for ${\tilde S}_x$ reads
\begin{equation}
\label{solution}
\hspace{-3mm}{\tilde S}_x=\frac{pp_1^2+p(\omega_L^z)^2+p_1\Omega_{\s SO}^2}{(pp_1)^2+2pp_1\Omega_{\s SO}^2+(p^2+p_1^2)(\omega_L^z)^2+[\Omega_{\s SO}^2-(\omega_L^z)^2]^2}.
\end{equation}
Four frequencies of the modes describing the spin dynamics are determined by the zeros of the denominator.
They are given by
 \begin{equation}
\label{frequencies}
p=\Omega_{\s SO}\Big[-\Gamma \pm i\Big(\frac{\omega_L^z}{\Omega_{\s SO}} \pm \sqrt{1-\Gamma^2}\Big)\Big],
\end{equation}
where the parameter $\Gamma$ is the dimensionless intervalley scattering rate  defined by Eq. (\ref{ratio}).

It is seen from Eq. (\ref{frequencies}) that the spin dynamics depends dramatically on the value of
$\Gamma$. For $\Gamma \ll 1$ there are two different oscillation frequencies,
$\Omega_{\s SO}\pm \omega_L^z$, which decay with the same rate, $\gamma_v$. On the contrary,
for $\Gamma \gg 1$ both frequencies are equal to $\omega_L^z$, but the decay rates are very different.
For the valley-symmetric mode it is equal to $2\gamma_v$, while the valley-antisymmetric mode decays
{\em very slowly} with the Dyakonov-Perel\cite{dyakonov-perel} rate $\approx \Omega_{\s SO}^2/2\gamma_v$.  The time evolution of $S_x(t)$
has different forms for $\Gamma <1$ and $\Gamma >1$. Namely, for $\Gamma < 1$ this evolution is given by
\begin{align}
\label{GammaSmall}
S_x(t)=&\frac{1}{2}\Bigg\{\frac{\Gamma}{\sqrt{1-\Gamma^2}}\Bigg[\sin \Bigg(\Big(\frac{\omega_L^z}{\Omega_{\s SO}} + \sqrt{1-\Gamma^2}\Big) \Omega_{\s SO}t\Bigg) \nonumber \\
- & \sin\Bigg(\Big(\frac{\omega_L^z}{\Omega_{\s SO}} - \sqrt{1-\Gamma^2}\Big)\Omega_{\s SO}t \Bigg)\Bigg]
 \nonumber \\
+& \Bigg[\cos\Bigg( \Big(\frac{\omega_L^z}{\Omega_{\s SO}}
+  \sqrt{1-\Gamma^2}\Big)\Omega_{\s SO}t\Bigg)
  \nonumber \\
+& \cos \Bigg(\Big(\frac{\omega_L^z}{\Omega_{\s SO}} - \sqrt{1-\Gamma^2}\Big)\Omega_{\s SO}t\Bigg)\Bigg]\Bigg\}\exp\Big[-\Gamma \Omega_{\s SO}t\Big],
 \end{align}
while for $\Gamma >1$ we have
\begin{align}
\label{GammaBig}
&S_x(t)=\frac{1}{2} \Bigg\{\Big(1+\frac{\Gamma}{\sqrt{\Gamma^2-1}}\Big)\exp\Big[\hspace{-1.2mm}-\hspace{-1.2mm}\Big(\Gamma-\sqrt{\Gamma^2-1}\Big) \Omega_{\s SO}t\Big] \nonumber \\
+&\Big(1-\frac{\Gamma}{\sqrt{\Gamma^2-1}}\Big)\exp\Big[-\Big(\Gamma+\sqrt{\Gamma^2-1}\Big) \Omega_{\s SO}t\Big]\Bigg\}\cos \omega_L^z t.
\end{align}

 \section{External field along ${\hat y}$}
For parallel magnetic field, there are, in general, six modes of the spin dynamics.
Although the spin dynamics in this geometry was considered in Ref. \onlinecite{Crooker1}, only the time evolution of $S_z$ was studied, while
we are interested in $S_x(t)$, $S_y(t)$. It turns out that the frequencies for $S_x(t)$ are the same
as for $S_z(t)$, while for  $S_y(t)$ they are completely different. This is certainly the specifics of
TMDs.

The field along ${\hat y}$ couples $S_x(t)$ and $S_z(t)$ via the conventional Larmor precession.
In addition, the valley-asymmetric field  $\pm{\hat z}\Omega_{\s SO}$ couples $S_x(t)$ to the spin {\em imbalance}, $S_y^-$. As a result, the system $6\times 6$ decouples into two systems $3\times 3$. The Laplace-transformed  system  involving ${\tilde S}_x$ reads
\begin{align}
\label{system2}
p{\tilde S}_x-1 &= -\Omega_{\s SO}{\tilde S}_y^- +\omega_L^y{\tilde S}_z \nonumber \\
p{\tilde S}_z &= -\omega_L^y{\tilde S}_x \nonumber \\
p_1{\tilde S}_y^- &= \Omega_{\s SO}{\tilde S}_x,
\end{align}
By contrast to the normal orientation, the solution
\begin{equation}
\label{ThirdOrder}
{\tilde S}_x=\frac{pp_1}{p^2p_1+p_1(\omega_L^y)^2 +p \Omega_{\s SO}^2 }
\end{equation}
contains a third-order polynomial in the denominator. With regard to sensitivity
of the spin dynamics to the external field, the most interesting case is $\Gamma \gg 1$,
when the intervalley scattering is fast. In this limit, the expressions for the two poles
have a simple form
\begin{equation}
\label{frequencies1}
p=\Omega_{\s SO}\Bigg[-\frac{1}{4\Gamma} \pm \sqrt{\Big(\frac{1}{4\Gamma}\Big)^2-\Big(\frac{\omega_L^y}{\Omega_{\s SO}}\Big)^2} \Bigg],
\end{equation}
and reproduce the corresponding frequencies obtained in Ref. \onlinecite{Crooker1}. Expression Eq. (\ref{frequencies1})
defines a small characteristic  magnetic field, $\omega_L^y\sim \Omega_{\s SO}/\Gamma$, which is the
inverse Dyakonov-Perel relaxation time. In the same limit, $\Gamma \gg 1$, the third frequency is given
by
 \begin{equation}
 \label{ThirdFrequency}
p=\Omega_{\s SO}\Bigg[-2\Gamma+\frac{2\Gamma}{4\Gamma^2+\Big(\frac{\omega_L^y}{\Omega_{\s SO}}\Big)^2} \Bigg].
\end{equation}
It corresponds to the decay with the rate $2\gamma_v$  and is insensitive to weak magnetic fields.

As magnetic field increases, the argument of the square root in Eq. (\ref{frequencies1}) changes sign.
This is reflected in the spin dynamics, which is different for $\omega_L^y$ bigger and smaller than
$\Omega_{\s SO}/4\Gamma$. At low fields we have
\begin{align}
\label{LowField}
S_x(t)=&\frac{1}{2} \Bigg[\Bigg(1+\frac{1}{\sqrt{1-\big(\frac{4\omega_L^y\Gamma}{\Omega_{\s SO}}\big)^2}}\Bigg)  \nonumber \\
\times & \exp \Bigg[-\Bigg(\frac{1}{4\Gamma} + \sqrt{\Big(\frac{1}{4\Gamma}\Big)^2-\Big(\frac{\omega_L^y}{\Omega_{\s SO}}\Big)^2} \Bigg)\Omega_{\s SO} t\Bigg]
\nonumber \\
+& \Bigg(1-\frac{1}{\sqrt{1-\big(\frac{4\omega_L^y\Gamma}{\Omega_{\s SO}}\big)^2}}\Bigg)  \nonumber \\
\times & \exp \Bigg[-\Bigg(\frac{1}{4\Gamma} - \sqrt{\Big(\frac{1}{4\Gamma}\Big)^2-\Big(\frac{\omega_L^y}{\Omega_{\s SO}}\Big)^2} \Bigg)\Omega_{\s SO} t\Bigg],
\end{align}
i.e. the dynamics is overdamped. It becomes oscillatory for $\omega_L^y>\Omega_{\s SO}/4\Gamma$. In this domain we find
\begin{align}
\label{HighField}
S_x(t)=&\exp\Big[-\frac{\Omega_{\s SO}}{4\Gamma}t\Big]\Bigg[\cos\Bigg(\sqrt{\Big(\frac{\omega_L^y}{\Omega_{\s SO}}\Big)^2-\Big(\frac{1}{4\Gamma}\Big)^2}\Omega_{\s SO} t\ \Bigg)
\nonumber \\
-&\frac{1}{\sqrt{\big(\frac{4\omega_L^y\Gamma}{\Omega_{\s SO}}\big)^2-1}}\sin\Bigg(\sqrt{\Big(\frac{\omega_L^y}{\Omega_{\s SO}}\Big)^2-\Big(\frac{1}{4\Gamma}\Big)^2}\Omega_{\s SO} t\ \Bigg)\Bigg].
\end{align}
Compared to Ref. \onlinecite{Crooker1}, where $S_z(t)$ was calculated, the amplitudes of the harmonics
in Eq. (\ref{HighField}) are different.

\begin{figure}
\includegraphics[width=84mm]{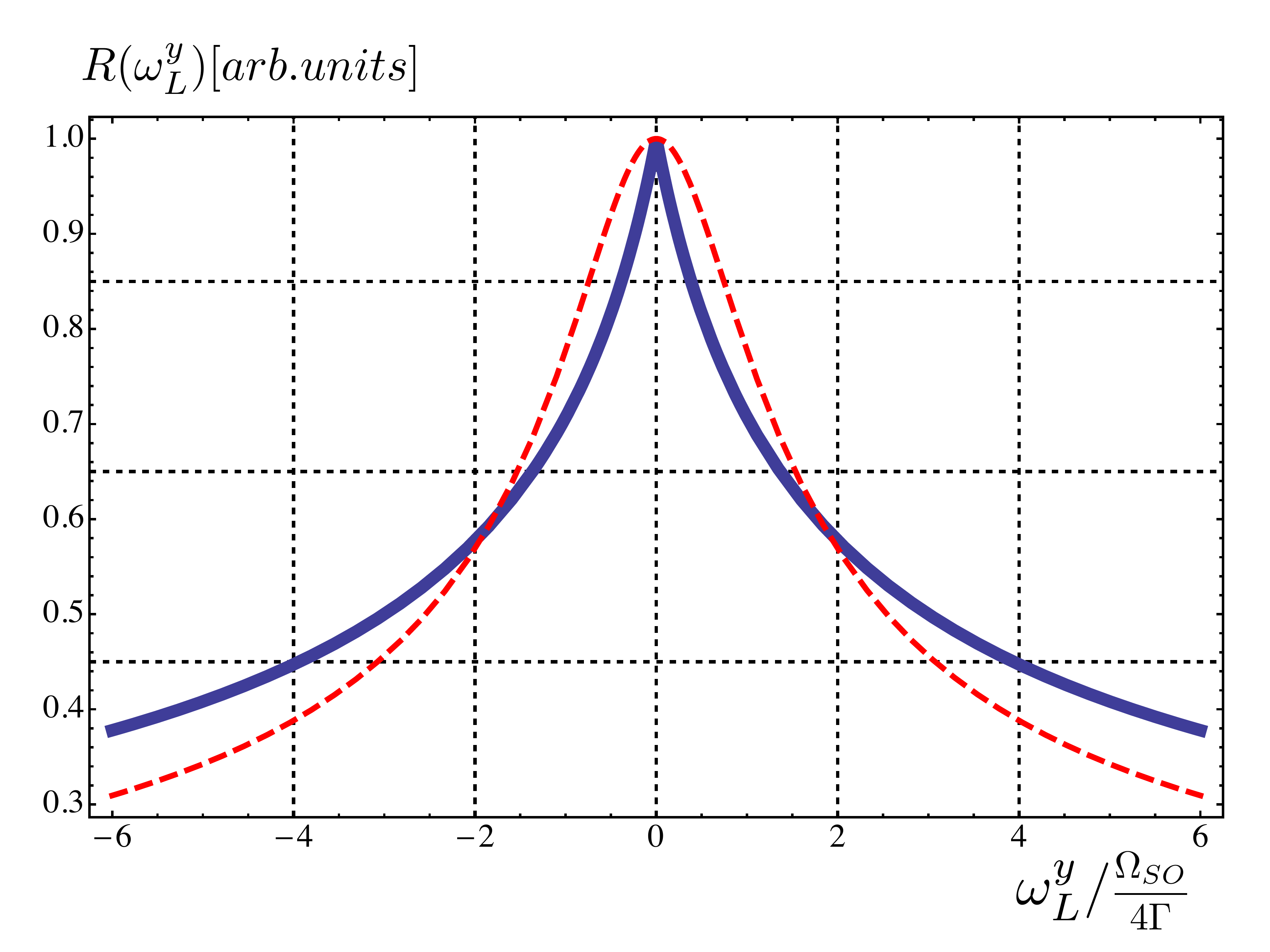}
\caption{(Color online) Universal shape of the Hanle curve, $R(4\omega_L^y\Gamma/\Omega_{\s SO})$, for the parallel orientation
of the magnetic field is plotted from Eq. (\ref{HanleParallel}). A cusp at zero field reflects the slow-decaying  valley-asymmetric mode
of the spin dynamics. For comparison, the conventional Hanle curve Eq. (\ref{G}) is plotted with dashed line.}
\label{figure3}
\end{figure}

\section{Shapes of the Hanle curves}
To find the Hanle profiles for normal orientation of magnetic field one
should substitute Eqs. (\ref{GammaSmall}) and (\ref{GammaBig}) into Eq. (\ref{R})
and perform the integration over time. The structure of $S_x(t)$, sinusoidal function times exponential
decay, suggests that  the integration can be carried out analytically\cite{Silsbee1985,Silsbee1988} for arbitrary ${\cal L}$ . However, in samples with low mobility, the regime of
small distance, ${\cal L}$, between  injector and detector is most relevant. This is because
the diffusion time, ${\cal L}^2/D$, should not exceed much the spin relaxation time.
Upon setting ${\cal L}=0$ in $P_{\cal L}(t)$ Eq. (\ref{propagator}) the integration is easily performed
with the help of the relations

\begin{align}
\label{smallL}
\int\limits_0^\infty \frac{dt~ e^{-vt}}{\sqrt{t}}\sin(ut)&= \text{sign}(u)\sqrt{\frac{\pi}{2v}}\mathrm{F}\Big(\frac{u}{v}\Big)
,\nonumber \\
\int\limits_0^\infty \frac{dt~e^{-vt}}{\sqrt{t}}\cos(ut)&= \sqrt{\frac{\pi}{2v}}\mathrm{G}\Big(\frac{u}{v}\Big),
\end{align}
where the functions $\mathrm{F}(z)$ and $\mathrm{G}(z)$ are defined as
\begin{equation}
\label{G}
\mathrm{F}\big(z \big)=\frac{\sqrt{\sqrt{1+z^2}-1}}{\sqrt{1+z^2}},
~~~~~\mathrm{G}\big(z \big)=\frac{\sqrt{\sqrt{1+z^2}+1}}{\sqrt{1+z^2}}.
\end{equation}
The expressions for nonlocal resistance in the $\perp$ geometry can be now expressed via the functions
$\mathrm{F}$ and $\mathrm{G}$. For $\Gamma <1$ we have
\begin{align}
\label{Gamma<1}
R(\omega_L^z)&=\frac{{\cal R}_{0}}{4\sqrt{2D \Omega_{\s SO} \Gamma}}
\nonumber \\ &\times
\Bigg\{\mathrm{G}\Big(\frac{\omega_L^z}{ \Omega_{\s SO}\Gamma}+\sqrt{\frac{1}{\Gamma^2}-1}\Big)+\mathrm{G}\Big(\frac{\omega_L^z}{ \Omega_{\s SO}\Gamma}-\sqrt{\frac{1}{\Gamma^2}-1}\Big)
\nonumber \\
&+ \frac{\Gamma}{\sqrt{1-\Gamma^2}}
\nonumber \\ &\times
\Bigg[\text{sign}\Big[\big(\frac{\omega_L^z}{ \Omega_{\s SO}}+\sqrt{1-\Gamma^2}\big) \Omega_{\s SO}\Big]\mathrm{F}\Big(\frac{\omega_L^z}{ \Omega_{\s SO}\Gamma}+\sqrt{\frac{1}{\Gamma^2}-1}\Big)
\nonumber \\
&-\text{sign}\Big[\big(\frac{\omega_L^z}{ \Omega_{\s SO}}-\sqrt{1-\Gamma^2}\big) \Omega_{\s SO}\Big]\mathrm{F}\Big(\frac{\omega_L^z}{ \Omega_{\s SO}\Gamma}-\sqrt{\frac{1}{\Gamma^2}-1}\Big)\Bigg]\Bigg\}.
\end{align}
The corresponding expression for $\Gamma >1$ reads

\begin{align}
\label{Gamma>1}
R(\omega_L^z)&=\frac{{\cal R}_{0}}{4\sqrt{2D \Omega_{\s SO} (\Gamma^2-1)}}
\nonumber \\
&\times\Bigg[\frac{1}{\Big(\Gamma-\sqrt{\Gamma^2-1}\Big)^{3/2}}\mathrm{G}
\Bigg(\frac{\omega_L^z}{ \Omega_{\s SO}}\Big(\Gamma+\sqrt{\Gamma^2-1}\Big)\Bigg)\nonumber \\
&-\frac{1}{\Big(\Gamma+\sqrt{\Gamma^2-1}\Big)^{3/2}}\mathrm{G}\Bigg(\frac{\omega_L^z}{ \Omega_{\s SO}}\Big(\Gamma-\sqrt{\Gamma^2-1}\Big)\Bigg)\Bigg].
\end{align}
Evolution of the shape of the Hanle curves with $\Gamma$ described by Eqs. (\ref{Gamma<1}), (\ref{Gamma<1}) is the following. For slow intervalley scattering $R(\omega_L^z)$ exhibits a two-peak
structure with maxima at $\omega_L^z\approx \pm \Omega_{\s SO}$. Each peak corresponds to the  ``compensation" of the SO-splitting in a given valley by the external field. The widths of the peaks are
$\sim \gamma_v$. For $\Gamma \approx 0.7$ the peaks merge, and, upon further increase of $\Gamma$,
transform into the {\em difference} of the two peaks with small, $\sim \Omega_{\s SO}/\Gamma$, and
 big, $\sim \Omega_{\s SO}\Gamma$, widths centered at $\Omega_L^z=0$. The broad peak, however, has a
 much smaller magnitude. So the shape for $\Gamma \gg 1$ is, essentially, the conventional Hanle shape
 with width determined by the inverse Dyakonov-Perel relaxation time. The evolution of $R(\omega_L^z)$
 with $\Gamma$ is illustrated in Fig. \ref{figure2}.

Turning to the geometry with external field along ${\hat y}$, we first observe that $S_x(t)$ given by
Eqs. (\ref{LowField}), (\ref{HighField}) contains only one scale of $\omega_L^y$, namely,
$\omega_L^y=\Omega_{\s SO}/\Gamma$. Since we assumed fast intervalley scattering, this characteristic field is much smaller than the splitting $\Omega_{\s SO}$. Naturally, the Hanle curve is a function of a single
parameter $\omega_L^y\Gamma/\Omega_{\s SO}$. The form of this function can be found using the identities
Eq. (\ref{smallL}). While the integrands in Eq. (\ref{R}) are different for $\omega_L^y<\Omega_{\s SO}/4\Gamma$ and $\omega_L^y>\Omega_{\s SO}/4\Gamma$, the resulting shape is given by a single concise expression
\begin{equation}
\label{HanleParallel}
R(\omega_L^y)=\frac{{\cal R}_{0}}{\sqrt{2D \Omega_{\s SO} }} \frac{\Gamma}
{\sqrt{1+\big\vert\frac{4\omega_L^z\Gamma}{\Omega_{\s SO}}\big\vert }}.
\end{equation}
The Hanle curve in the form of Eq. (\ref{HanleParallel}) falls off with magnetic field
as $1/\sqrt{\omega_L^y}$, i.e.  in the same way  as a regular Hanle curve. However, it
exhibits a unique feature at small field, where the slope has an abrupt cusp, see Fig. \ref{figure3}.
The origin of the cusp can be traced to the second term in Eq. (\ref{LowField}). Rather than spin precession, this term describes the slow decay, as $\exp(-\vert\omega_L^y\vert t)$, of the spin density. This slow decay reveals the specifics of the two-valley spin dynamics Eq. (\ref{system2}), for which the
valley-asymmetric mode decays anomalously slow.

\section{Concluding remarks}
\noindent{(i)} In optics experiments the valleys were ``addressed" separately, in the sense,
that, for a given frequency, different polarizations of light generated excitons in different valleys.
Conventional spin transport is valley-insensitive. On the other hand, the spin-pumping setup\cite{Tserkovnyak} can
serve as an analog of optical selective valley excitation. Assume that the ferromagnetic resonance
is excited in the  ferromagnet which injects spin into a TMD layer. The pumped spin current would flow
in one of the valleys depending on polarization of the microwave field exciting the resonance.
The analogy between the selective valley excitation in optics and by the spin pumping is straightforward. While in optical experiments an absorbed photon generates an electron in the conduction band and a hole in the valence band, in spin pumping it is a magnon that creates an  electron-hole pair
 in the Fermi sea of the conduction band.

\noindent{(ii)}
For fast intervalley scattering, the slow-decaying modes of the spin dynamics are present for  both orientations of the external field. These modes are valley-asymmetric and originate from almost complete
compensation of the SO field, $\Omega_{\s SO}$, in the valley $K$ and $-\Omega_{\s SO}$
in the valley $K'$. However, the decay rates of these modes are drastically different in weak external
field. This is because, for $\perp$ orientation, the result of compensation is simply the external
field ${\hat z}\omega_L^z$, while, for $\parallel$ orientation, the result of compensation of
${\hat y}\omega_L^y+{\hat z}\Omega_{\s SO}$ and  ${\hat y}\omega_L^y-{\hat z}\Omega_{\s SO}$
is only quadratic in external field. Then the prime effect of the external field on the spin-dynamics is the field-dependent decay.

\noindent{(iii)}. Measuring the Hanle curves in both $\parallel$ and $\perp$ orientations
allows, in principle, to determine the values of both relevant parameters, $\Omega_{\s SO}$ and $\gamma_v$.

\noindent{(iv)}. Our main results Eqs. (\ref{Gamma<1}), Eq. (\ref{Gamma>1}), and Eq. (\ref{HanleParallel}
were derived in the limit of small distance ${\cal L}$ between injector and detector. Now we can quantify
the corresponding condition. Characteristic magnetic field in  Eq. (\ref{HanleParallel}) is
$\omega_L^y\sim \Omega_{\s SO}/\Gamma$. Thus, the diffusion time, ${\cal L}^2/D$ should be smaller than the
precession time, i.e. ${\cal L}\ll \Big(D\Gamma/\Omega_{\s SO}\Big)^{1/2}$. In the opposite limit the Hanle
curve exhibits sensitivity to even weaker fields. The corresponding expression for nonlocal resistance in this limit
can be cast in the form

\begin{equation}
\label{last}
\frac{R(\omega_L^y)-R(0)}{R(0)}=-\frac{2|\omega_L^y|\Gamma}{\Omega_{\s SO}}\exp\Big[-2|\omega_L^y|\Big(\frac{{\cal L}^2\Gamma}{D\Omega_{\s SO}}\Big)^{\frac{1}{2}}\Big].
\end{equation}
This result suggests that the Hanle curve has a minimum at zero field and two maxima at $\omega_L^y=\pm \Big(D\Omega_{\s SO}/4{\cal L}^2\Gamma\Big)^{1/2}$ much smaller than $\Omega_{\s SO}/\Gamma$.

\noindent{(v)} In experimental paper Ref. \onlinecite{Crooker2} the sensitivity of the optical
response to the normal magnetic field was not registered until $\omega_L^z$ was as high as $65$T.
We, on the other hand, predict the sensitivity of the spin transport to much weaker fields. The reason
is that $z$-projection of spin, $S_z(t)$, registered in optical experiments, drops out from the equations Eq. (\ref{system1}) for the spin dynamics in $\perp$ orientation, whereas the dynamics, $S_x(t)$, relevant
for spin transport, persists.

\noindent{(vi)} Hanle shapes with minima at zero external field, like  Fig. \ref{figure2} for $\perp$ orientation and
Eq. (\ref{last}) for $\parallel$ orientation are unique and constitute our main verifiable  prediction. More experimental studies of non-local Hanle measurements with non-optical spin injection techniques (preferably electrical spin injection) are needed to fully understand the peculiar spin transport characteristics of TMD films.

\section{Acknowledgements}

 This work was supported by NSF through MRSEC DMR-1121252.


\begin{thebibliography}{99}
\bibitem{footnote2}
We assume that the Fermi level is high, so that this splitting is much bigger the intrinsic splitting.\cite{KP4}
\bibitem{footnote1}
We neglected the intraband spin relaxation  introduced phenomenologically in Ref. \onlinecite{Crooker1}
since it does not lead to the qualitatively new effects.
\end{thebibliography}
\end{document}